\documentclass[aps,prl,twocolumn,superscriptaddress,groupedaddress]{revtex4}  
\usepackage{graphicx}
\usepackage{amsmath,amssymb}
\usepackage{float}
\usepackage{color}

\newcommand {\NF}{{N_{\rm F}}}

\newcommand {\bbk}{{\bf k}}
\newcommand {\bbq}{{\bf q}}
\newcommand {\bbkk}{{\bf {k'}}}
\newcommand {\ek}{{\epsilon_{\bf k}}}

\newcommand {\oql}{{\omega_{\bf q \nu}}}
\newcommand {\on}{{\omega_n}}
\newcommand {\onp}{{\omega_{n'}}}
\newcommand {\afo}{{\alpha^{2} F(\omega)}}
\newcommand {\gkkl}{{g_{\bf k \bf {k'}}^{\nu}}}
\newcommand {\deltak}{{\delta(\epsilon_{\bf k})}}
\newcommand {\deltakk}{{\delta(\epsilon_{\bf {k'}})}}

\begin{document}
\title{First-principles calculations of the superconducting properties in Li-decorated monolayer graphene within the anisotropic Migdal-Eliashberg formalism}

\author{Jing-Jing Zheng}
\affiliation{Department of Physics, Applied Physics and Astronomy,
Binghamton University-SUNY, Binghamton, New York 13902, USA}
\affiliation{Institute of Theoretical Physics and Department of
Physics, Shanxi University, Taiyuan 030006, People's Republic of
China}
\author{E. R. Margine }
\affiliation{Department of Physics, Applied Physics and Astronomy,
Binghamton University-SUNY, Binghamton, New York 13902, USA}

\begin{abstract}

The \emph{ab initio} anisotropic Migdal-Eliashberg formalism has been used
to examine the pairing mechanism and the nature of the
superconducting gap in the recently discovered lithium-decorated
monolayer graphene superconductor. Our results provide evidence that
the  superconducting transition in Li-decorated monolayer graphene
can be explained within a standard phonon-mediated mechanism. We
predict a single anisotropic superconducting gap and a critical
temperature $T_c=$ 5.1-7.6~K, in very good agreement with the
experimental results.
\end{abstract}
\maketitle

\section{INTRODUCTION}
During the past decade graphene has revolutionized many areas of
nanotechnology from organic electronics to photovoltaics,
plasmonics, photonics, and energy storage~\cite{Novoselov_NAT12}.
One notable application that was missing from this list was
superconductivity, despite numerous theoretical predictions of
either a conventional or an unconventional pairing
mechanism~\cite{einenkel,profeta,Margine2014,uchoa,nandkishore,kiesel}.
Very recently, a high-resolution angle-resolved photoemission
spectroscopy (ARPES) study has presented evidence supporting the
appearance of a superconducting phase in Li-decorated monolayer
graphene (LiC$_6$) around 5.9~K~\cite{Ludbrook}, within the standard
phonon-mediated coupling mechanism. This work has been followed by
two more studies that reported the observation of superconductivity
in Ca-intercalated bilayer graphene~\cite{Hasegawa} and in
Ca-intercalated graphene laminates~\cite{Chapman}.

In this article, we investigate from first principles the nature of
the superconducting gap in LiC$_6$. To this end, we solve the fully
anisotropic Migdal-Eliashberg
equations~\cite{allen_mitrovic,Margine2013} to obtain the
superconducting transition temperature ($T_c$) and the variation of
the superconducting energy gap on the Fermi surface. While previous
\emph{ab initio} calculations have shown that the electron-phonon
coupling is sufficient to yield a critical temperature in 6.7-10.3~K
range using the Allen-Dynes formula~\cite{profeta,Guzman,kaloni} or
the isotropic Eliashberg formalism~\cite{Szcz2014}, the nature of
the superconducting gap has not yet been addressed. We find that,
similar to bulk
CaC$_{6}$~\cite{Sanna_PRB07,Margine2016,Szcz2014,Szcz2015},
Li-decorated monolayer graphene exhibits a single anisotropic gap in
agreement with the experimental work~\cite{Ludbrook}.

\begin{figure}[ptb]
\begin{center}
\includegraphics[width=\linewidth]{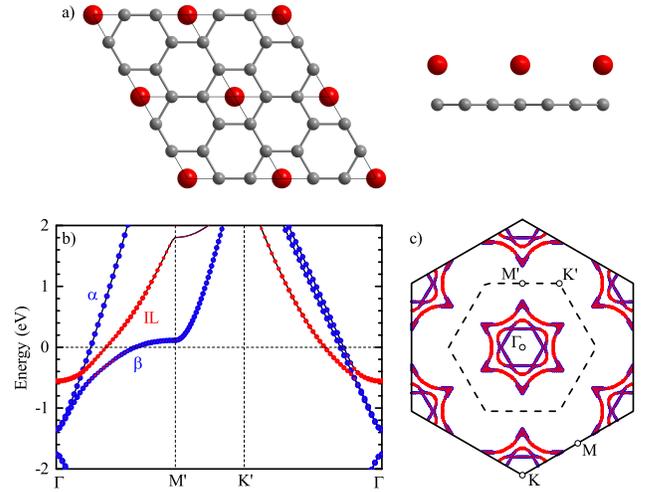}
\end{center}
\caption{(Color online)(a) Top- and side-view of a ball-and- stick
model of LiC$_6$, with C in gray and Li in red. (b) Band structure
of LiC$_6$. The inner and outer $\pi^*$ bands (with respect to
$\Gamma$ point) are labeled as $\alpha$ and $\beta$ (blue dots). The
interlayer band is labeled as IL (red dots). The size of the blue
and red symbols is proportional to the contribution of
C-$\emph{p$_{z}$}$ and Li-$\emph{s}$ character. (c) The
two-dimensional Fermi surface of LiC$_{6}$ with the same color code
as in (b). The Brillouin zones of a graphene unit cell and a
$\sqrt{3} \times \sqrt{3} R$30$^\circ$ graphene supercell are shown
as black full and dashed lines, respectively.
}%
\label{fig1}%
\end{figure}

\begin{figure*}[t]
\begin{center}
\includegraphics[width=150mm]{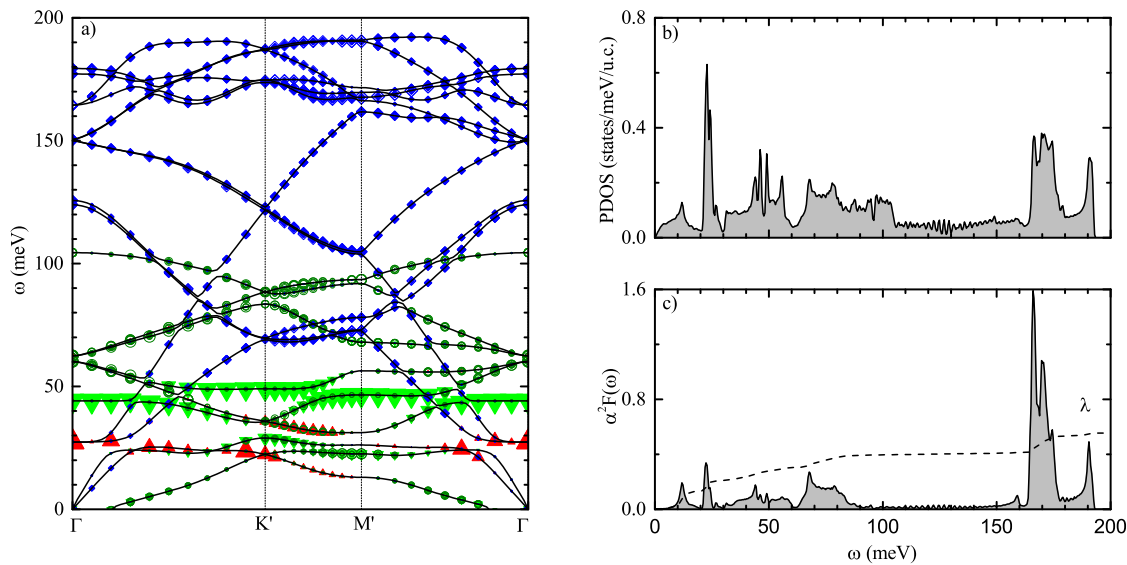}
\end{center}
\caption{(Color online) (a) Phonon frequency dispersion of LiC$_{6}$. The decomposition of the phonon
spectrum with respect to C and Li atomic vibrations is indicated by: olive circle (C$_{z}$),
blue diamond (C$_{xy}$), red triangle up (Li$_{xy}$), and green triangle down (Li$_{z}$). (b) Phonon density of states  and (c) Eliashberg spectral function with cumulative electron-phonon coupling strength of LiC$_6$. The solid line is for $\alpha^{2}F(\omega)$, the dashed line is for $\lambda(\omega)$.}%
\label{fig2}%
\end{figure*}

\section{METHODOLOGY}
The calculations are performed within the local density
approximation to density-functional theory~\cite{lda1,lda2} using
planewaves and norm-conserving pseudopotentials~\cite{nc1,nc2}, as
implemented in the {\tt Quantum-ESPRESSO} package~\cite{QE}. The
planewaves kinetic energy cutoff is 100~Ry and the structural
optimization is performed until the forces on atoms are less than
10~meV/\AA\,. Li-decorated monolayer graphene is described in the
$\sqrt{3} \times \sqrt{3} R$30$^\circ$ graphene supercell with one
lithium atom per unit cell. The optimized lattice constant and the
adatom-graphene distance are $a=4.24$~\AA\, and $h=1.78$~\AA. A
Brillouin-zone (BZ) $\Gamma-$centered $\mathbf{k}$-point mesh of
24$\times$24 and a Methfessel-Paxton smearing~\cite{mp} of 0.02~Ry
are adopted for the electronic charge density calculations. The
phonon modes are computed within density-functional perturbation
theory~\cite{baroni2001} on a 6$\times$6 $\mathbf{q}$-mesh. We
employ the EPW
code~\cite{Giustino2007,Giustino2016,EPW,Margine2013,Ponce} to
obtain the superconducting gap. The calculation of the electronic
wavefunctions required for the Wannier-Fourier
interpolation~\cite{Marzari,Mostofi} in EPW is performed on a
uniform unshifted BZ $\mathbf{k}$-point grid of size 12$\times$12.
For the anisotropic Eliashberg equations, we use 120$\times$120 and
60$\times$60 $\mathbf{k}$- and $\mathbf{q}$-point grids. The
Matsubara frequency cutoff is set to five times the largest phonon
frequency ($5\times 200$~meV), and the Dirac delta functions are
replaced by Lorentzians of widths 100~meV and 0.5~meV for electrons
and phonons, respectively.

\section{ELECTRONIC AND VIBRATIONAL PROPERTIES}
In Figs.~\ref{fig1} (a)-(c), we show the crystal structure of
Li-decorated monolayer graphene along with the corresponding
decomposed electronic band structure and Fermi surface. Three bands
cross the Fermi level around the $\Gamma$ point, in agreement with
previous reports~\cite{profeta,Guzman}. The inner and outer C
$\pi^*$ bands, labeled as $\alpha$ and $\beta$ (blue dots), are
obtained by folding the $\pi^*$ states of graphene from $K$ to
$\Gamma$, following the superstructure induced by Li adsorption. The
Li-derived band, labeled as IL (red dots), displays a nearly-free
electron like dispersion upwards from about 0.56~eV below the Fermi
energy. Similar weakly bound free-electron states have been observed
in other layered materials~\cite{profeta,Calandra,Boeri,Kolmogorov}
and nanotubes~\cite{Margine_NFE}, and their rapid downshift under
doping is due to the combined effects of quantum confinement and
electrostatic response~\cite{profeta,Boeri,Margine_NFE}. The
corresponding Fermi surface of LiC$_6$ can be divided into two
concentric regions centered around the $\Gamma$ point. The inner
region is characterized by a snowflake-like electron pocket
intersecting a hexagonal electron pocket, which arises from the
mixing of the inner C $\pi^*$ states with the Li $s$ states. These
Fermi sheets resemble the $\Gamma$-centered Fermi surface observed
in Ca and Li intercalated bilayer
graphene~\cite{Margine2016,Shimizu_PRL14}. The outer region also has
a snowflake-like shape and originates on the outer C $\pi^*$ states
and the Li $s$ states.

We now focus on the vibrational properties and the electron-phonon
coupling (EPC) in LiC$_{6}$. Similar to bulk CaC$_6$~\cite{Calandra}
and bilayer C$_6$CaC$_6$~\cite{Margine2016}, one can clearly
identify in Fig.~\ref{fig2}(a) three regions in the phonon
dispersion associated to (i) the Li-related modes (up to 50~meV,
where above 37~meV are Li$_z$ modes mixed with carbon out-of-plane
C$_z$ modes), (ii) the carbon out-of-plane C$_z$ vibrations
(50-100~meV), and (iii) the carbon in-plane C$_{xy}$ modes (above
100~meV). The size of the symbols in Fig.~\ref{fig2}(a) is
proportional to the atomic displacements corresponding to Li and C
in-plane and out-of-plane contributions.

The isotropic Eliashberg spectral function $\alpha^{2}F(\omega)$
\begin{equation}
\afo = \frac{1}{\NF N_{\bf k} N_{\bf q}} \sum_{\bbk,\bbkk,\nu} |\gkkl|^2 \deltak \deltakk \delta(\omega-\oql),
\end{equation}
and the cumulative electron-phonon coupling strength $\lambda(\omega)$
\begin{equation}
\lambda(\omega) = 2 \int_{0}^{\omega} d\omega' \alpha^2 F(\omega')/\omega',
\vspace{-0.1cm}
\end{equation}
are shown in Figs.~\ref{fig2}(b)-(c). In these expressions $\NF$
represents the density of electronic states per spin at the Fermi
level, $N_{\bf k}$ and $N_{\bf q}$ are the total numbers of ${\bf
k}$ and ${\bf q}$ points, $\ek$ is the Kohn-Sham eigenvalue with
respect to the Fermi level, and $\gkkl$ is the screened
electron-phonon matrix element for the scattering between the
electronic states $\bbk$ and $\bbkk$ through a phonon with wave
vector $\bbq\!=\!\bbkk\!-\bbk$, frequency $\oql$ and branch
index~$\nu$. Here $\bbk$ and $\bbkk$ indicate both the electron
wavevector and the band index. We find that the low-energy phonons
are key to achieving a high electron-phonon coupling in LiC$_6$ as
they account for 0.28 (51\%) of the total EPC ($\lambda$ = 0.55). On
the other hand, the electron-phonon coupling strengths associated
with the out-of-plane C$_z$ and in-plane C$_{xy}$ modes are 0.12
(22\%) and 0.15 (27\%), respectively. Such behavior has also been
found in bilayer C$_6$CaC$_6$, where the most significant
contribution to the EPC comes from the low-energy phonon
modes~\cite{Margine2016}. Overall, our calculated EPC $\lambda=0.55$
is in good agreement with the experimental value $0.58\pm0.05$
observed at the highest Li coverage~\cite{Ludbrook} and the values
reported in previous theoretical studies~\cite{profeta,Guzman}.

\begin{figure}[ptb]
\begin{center}
\includegraphics[width=75mm]{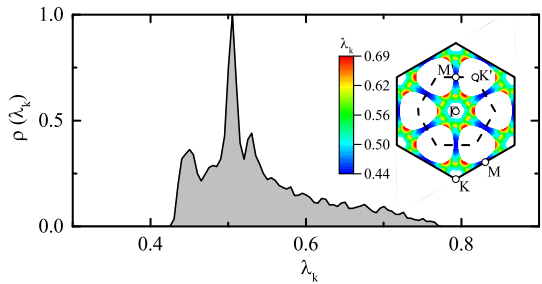}
\end{center}
\caption{(Color online) Distribution of the electron-phonon coupling strength $\lambda_{\mathbf{k}}$ of LiC$_6$. Inset: Momentum-resolved electron-phonon coupling parameters $\lambda_{\mathbf{k}}$ on the Fermi surface (the data points correspond to electrons within $\pm 150$~meV from the Fermi energy).}%
\label{fig3}%
\end{figure}

\begin{figure*}[t]
\begin{center}
\includegraphics[width=150mm]{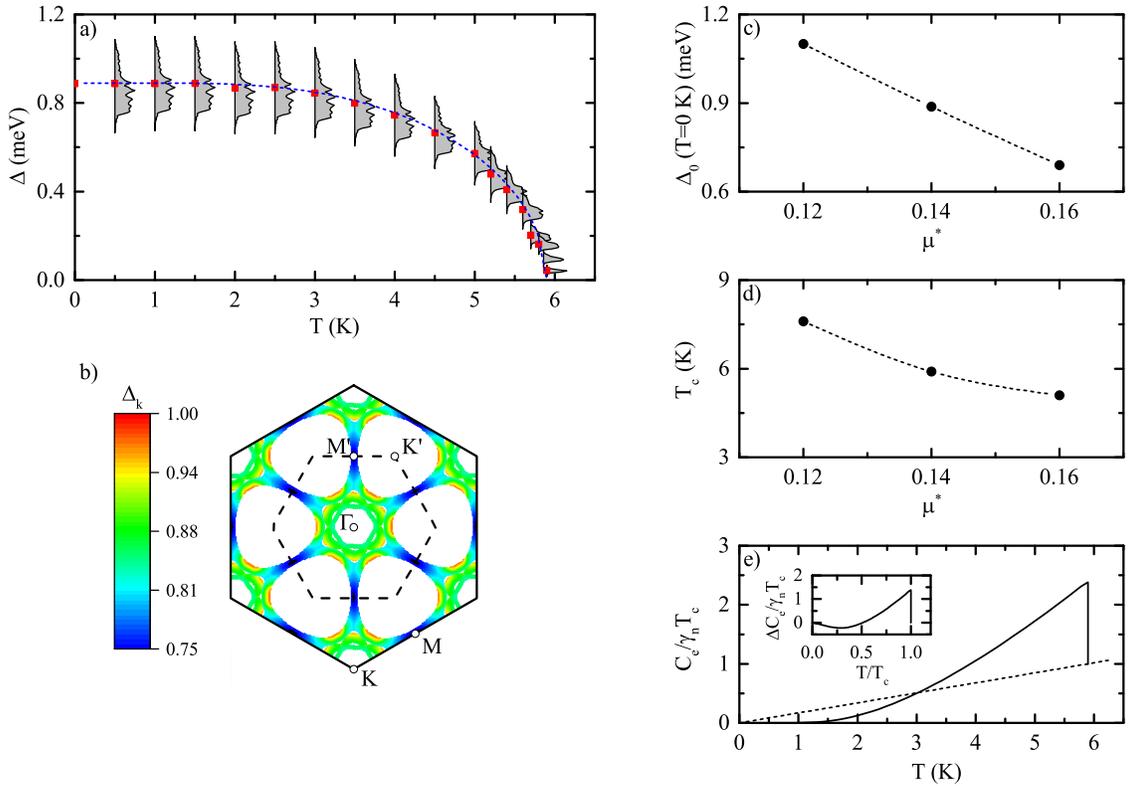}
\end{center}
\caption{(Color online) (a) Energy distribution of the anisotropic
superconducting gap $\Delta_{\mathbf{k}}$ of LiC$_{6}$ as a function
of temperature. The gap was calculated using a Coulomb
pseudopotential $\mu^{*}$ of 0.14. The red squares represent the
average value of the gap which vanishes at the critical temperature
\emph{$T_{c}$} = 5.9~K. The blue dashed line is the BCS fit to the
calculated data. (b) Momentum-resolved superconducting gap
$\Delta_{\mathbf{k}}$ (in meV) on the Fermi surface at 0.5~K. The
data points correspond to electrons within $\pm 150$~meV from the
Fermi energy. (c) Calculated superconducting gap at the Fermi level
in the \emph{T} = 0~K limit as a function of the Coulomb parameter
$\mu^{*}$. (d) Calculated superconducting critical temperature as a
function of the Coulomb parameter $\mu^{*}$. (e)
Normalized specific heat as a function of temperature for $\alpha$ = 1.75 in the superconducting state (solid line) and normal state (dashed line). The specific heat difference between the normal and superconducting state versus reduced temperature is shown in the inset.}%
\label{fig4}%
\end{figure*}

To quantify the anisotropy in the electron-phonon coupling, we
further evaluate the momentum-resolved EPC
$\lambda_\mathbf{k}$~\cite{Margine2013}, defined as:
\begin{equation}
\lambda_{\bbk} = \sum_{\bbkk,\nu} \deltakk |\gkkl|^2/\oql.
\vspace{-0.1cm}
\end{equation}
The calculated $\lambda_\mathbf{k}$ displays a significant
anisotropy with a distribution in the 0.42-0.78 range as shown in
Fig.~\ref{fig3}. This is in line with experimental ARPES
measurements where a marked anisotropy in the electron-phonon
coupling has been observed in the case of decorated
graphene~\cite{Ludbrook,Fedorov_NatCom14}, intercalated bilayer
graphene~\cite{Kleeman_JPSJ14}, and intercalated
graphite~\cite{Gruneis_PRB09,Valla_PRL09}. An alternative way to
look at the EPC anisotropy is presented in the inset of
Fig.~\ref{fig3}, where the variation of $\lambda_\mathbf{k}$ on the
Fermi surface is shown. When compared with the Fermi surface plot in
Fig.~\ref{fig1}(c), one can clearly see that the largest value of
$\lambda_\mathbf{k}$ is attained on the portions of the Fermi
surface dominated by the Li states. Notably, in bulk CaC$_6$,
$\lambda_\mathbf{k}$ was also found to be larger for the states with
Ca dominant orbital character on the Fermi
surface~\cite{Sanna_PRB07}. An important implication of this finding
is that the IL state and its associated interaction play a critical
role in the superconducting paring of LiC$_6$. A recent ARPES study
has provided compelling evidence regarding the importance of the IL
band in the pairing mechanism of bulk CaC$_6$~\cite{Yang}.
Furthermore, the lack of any sign of superconductivity down to 3.5~K
in few-layer graphene under large charge doping induced by
electrochemical gating~\cite{Gonnelli} provides additional proof of
the vital role of dopant atoms.

\section{SUPERCONDUCTING PROPERTIES}
The superconducting properties of LiC$_6$ are obtained by solving
self-consistently the fully anisotropic Migdal-Eliashberg equations
along the imaginary axis at the fermion Matsubara frequencies
$\on=(2n+1)\pi T$ (with $n$ an integer) for each temperature $T$
~\cite{Margine2013,Margine2014,allen_mitrovic,Choi}:
\begin{eqnarray} \hspace{-0.25cm}
  &&Z(\bbk,i\on) =
   1 + \frac{\pi T}{\NF \on} \sum_{\bbkk n'}
   \frac{ \onp }{ \sqrt{\onp^2+\Delta^2(\bbkk,i\onp)} } \nonumber \\
  &&\qquad\qquad\qquad\times \deltakk \lambda(\bbk,\bbkk,n\!-\!n'),
  \label{Znorm_surf} \\
 &&Z(\bbk,i\on)\Delta(\bbk,i\on) =
   \frac{\pi T}{\NF} \sum_{\bbkk n'}
   \frac{ \Delta(\bbkk,i\onp) }{ \sqrt{\onp^2+\Delta^2(\bbkk,i\onp)} } \nonumber \\
  &&\qquad\qquad\qquad\times \deltakk \left[ \lambda(\bbk,\bbkk,\!n-\!n')-\mu_c^*\right].\nonumber\\
\label{Delta_surf}
\end{eqnarray}
$Z(\bbk,i\on)$ is
the mass renormalization function, $\Delta(\bbk,i\on)$ is the
superconducting gap function, $\lambda(\bbk,\bbkk,\!n-\!n')$ is the
momentum- and energy-dependent EPC, and $\mu_c^*$ is the
semiempirical Coulomb parameter. The anisotropic $\lambda(\bbk,\bbkk,\!n-\!n')$ to be used in the Eliashberg equations is given by:
\begin{equation} \label{lambda}
\lambda(\bbk,\bbkk,n - n') = \NF \sum_{\nu}
\frac{2\omega_{\bbq \nu}}{(\on - \onp)^2+\omega_{\bbq \nu}^2} |\gkkl|^2.
\end{equation}

Figure~\ref{fig4}(a) shows the superconducting energy gap
$\Delta_\mathbf{k}$ as a function of temperature, calculated for a
screened Coulomb parameter $\mu^* = 0.14$, together with the average
value of the gap (red squares). The superconducting gap
$\Delta_\mathbf{k}$ on different parts of the Fermi surface at 0.5~K
is shown in Fig.~\ref{fig4}(b). We find that monolayer LiC$_{6}$
displays a single anisotropic gap with an average value $\Delta_0 =
0.89$~meV in the $T=0$~K limit, in very good agreement with the
ARPES result of $0.9 \pm 0.2$~meV, measured at
3.5~K~\cite{Ludbrook}. This situation is similar to bulk CaC$_{6}$
where the multiple-sheet Fermi surface gives rise to a single gap
structure with a sizable anisotropy~\cite{Sanna_PRB07,Margine2016},
but unlike bilayer C$_{6}$CaC$_{6}$ for which a two gap structure
has been recently predicted~\cite{Margine2016}.

The superconducting $T_{c}$ is identified as the highest temperature
at which the gap vanishes. From Fig.~\ref{fig4}(a) we find $T_{c} =
5.9$~K and a ratio $2 \Delta_0 / k_B T_c = 3.50$, very close to the
ideal BCS value of 3.53~\cite{BCS}. The predicted superconducting
critical temperature is in excellent agreement with the experimental
estimation of 5.9~K based on measurements of the size of the
superconducting gap~\cite{Ludbrook}. The temperature dependence of
the superconducting gap can be well fitted with a BCS model, as
obtained by solving numerically the BCS gap equation~\cite{Johnston}
with $\Delta_0$ and $T_c$ from our first-principles calculations.
This is shown by the blue dashed line in Fig.~\ref{fig4}(a). These
results provide support for a conventional phonon-mediated mechanism
as the superconducting origin in Li-decorated graphene. For
completeness, we also explore the sensitivity of the calculated
superconducting energy gap and critical temperature to the choice of
the Coulomb parameter $\mu^*$, as shown in Figs.~\ref{fig4}(c)-(d).
For $\mu^*$ = 0.12 and 0.16, we obtain $\Delta_0$ = 1.10~meV and
0.69~meV and $T_{c}$ = 7.6 and 5.1~K, respectively.

Finally, using the $\alpha$--model~\cite{Johnston,Padamsee}, we
obtain the temperature dependence of the reduced electronic specific
heat in the superconducting state. Within this model, the ratio
$\alpha$ = $\Delta_0 / k_B T_c$ is an adjustable parameter and the
normalized superconducting state electronic entropy $S_{es}$ and
heat capacity $C_{es}$ are  expressed in terms of $\gamma_n T_c$ as:
\begin{equation} \label{entropy}
\frac{S_{es}(t)}{\gamma_{n}T_{c}} =-
\frac{6\alpha}{\pi^{2}}\int^{\infty}_{0}[f\ln(f)+(1-f)\ln(1-f)]d \tilde{\varepsilon},
\end{equation}
\begin{equation} \label{specificheat}
\frac{C_{es}(t)}{\gamma_{n}T_{c}} =
 t\frac{d(S_{es}/\gamma_{n}T_{c})}{dt},
\end{equation}
where $f= \left[\exp(\alpha \tilde{E}/t)+1\right]^{-1}$ is the Fermi-Dirac distribution function, $t=T/T_c$ is the reduced temperature, and $\gamma_n =(2/3) \pi^2 k_B^2 \NF$ is the Sommerfeld coefficient.
The reduced quasi-particle energy is defined as $\tilde{E}=\sqrt{\tilde{\varepsilon}^{2}+\delta^{2}(t)}$, where $\tilde{\varepsilon}= \varepsilon/\Delta_{0}$ is the reduced
normal state single-particle energy relative to the Fermi level and $\delta(t) =
\Delta(T)/\Delta_{0}$ is the reduced gap function. The upper limit in the integral in Eq.~(\ref{entropy}) is set to $500 \gg 1$.

Figure~\ref{fig4}(e) shows the calculated $C_{es}/\gamma_n T_c$ for
$\alpha=1.75$. The temperature dependence of the normalized gap
$\delta(t)$ is assumed to be the same as in the BCS
theory~\cite{Johnston,Padamsee}. We checked the numerical results by
comparing the data for $\alpha_{BCS}=1.764$ with Tables II--IV in
Ref.~[\onlinecite{Johnston}] and by verifying that the entropy at
the critical temperature is equal to that of the normal state. The
shape of the calculated specific heat curve is consistent with a
one-gap BCS model and undergoes a discontinuous jump at the critical
temperature. The specific heat jump at $T_c$ is found to be $\Delta
C_e (T_c) / \gamma_n T_c = 1.385$ [shown in the inset of
Fig.~\ref{fig4}(e)], close to the weak limit BCS value of
1.426~\cite{BCS}. Furthermore, this result is comparable to the
experimetal and theoretical values reported for the normalized
specific heat jump in bulk CaC$_6$~\cite{Kim,Sanna_PRB07}.

\section{CONCLUSIONS}
In conclusion, we have studied the superconducting properties in Li-decorated
monolayer graphene within the {\it ab initio} anisotropic Migdal-Eliashberg theory. Our results provide support for a standard phonon-mediated
mechanism at the origin of the superconducting transition.  Most of the electron-phonon
coupling originates from the low-energy modes dominated by the motion of Li atoms
similar to bilayer C$_6$CaC$_6$. We find a sizable anisotropy in the electron-phonon
coupling which yields a single anisotropic gap over the Fermi surface.  Further enhancement in the critical temperature of LiC$_6$ is expected in the presence of a substrate~\cite{kaloni} or under applied strain~\cite{Pesic2014}.

\section{ACKNOWLEDGMENTS}
J-J Zheng acknowledges the support from the China Scholarship Council
(Grant No. 201508140043).

\end{document}